# *Mm-wave specific challenges in designing 5G transceiver architectures and air-interfaces*

Mythri Hunukumbure[1], Jian Luo[2], Mario Castañeda[2], Raffaele D'Errico[3], Per Zetterberg[4], Ali A. Zaidi[5], Jaakko Vihriälä[6], Domenico Giustiniano[7]

[1]*Samsung Electronics UK*, [2]*Huawei Technologies Duesseldorf GmbH*, [3]*CEA-LETI*, [4]*Qamcom AB*, [5]*Ericsson AB*, [6]*Nokia*, [7]*IMDEA Networks Institute*

*Abstract*—The mm-wave spectrum will be of significant importance to 5G mobile systems. There are multiple challenges in designing transceiver architectures and air interfaces in this spectrum. This paper is an attempt to explain some of these challenges and their interactions as means of enabling robust system design in near future.

*Keywords—air interface; hardware impairments; hybrid beamforming; mm-wave; waveforms;*

## I. INTRODUCTION

The 5[th] Generation of mobile communications (5G) is envisioned to offer extreme broadband, massive connectivity and very low latency with ultra-high reliability [1]. It is generally acknowledged that these challenging demands can only be met with a significant increase in the available bandwidths for 5G systems. The sub-6 GHz bandwidths are already heavily used and fragmented, making the spectrum above 6 GHz very appealing for 5G operations. Within the 6-100 GHz spectrum (loosely defined as mm-wave here), there are large swathes of spectrum where mobile is identified as a co-primary user [2]. The World Radio-communications Conference (WRC) 2015 identified several of the mm-wave bands for further study for mobile applications [3], thus raising the prospects for future worldwide use of these spectrum.

Proper mobile radio system design in these frequencies would enable the mm-wave spectrum to be effectively utilized for 5G operations. The EU-H2020 funded mm-MAGIC project [4] aims to design and develop Radio Access Technologies (RAT) in this mm-wave spectrum, to meet the challenging KPIs (Key Performance Indicators) of the identified 5G use cases. Within the RAT design, the transceiver architecture and the air-interface design are important aspects, which are discussed in this paper. We will specifically look at the challenges inherent to the mm-wave spectrum, which will be addressed by the solutions designed later in the project.

The need for multi-antenna beamforming architectures is already well recognized to mitigate the high path loss (as per Friis law) experienced in the mm-wave spectrum. However, this type of directional communications brings in a lot of challenges for the transceiver architecture as well as air interface design. In this paper, we will discuss the transceiver architectural options and challenges in section II. In terms of beamforming, the analysis on architectures to provide the best trade-offs in terms of performance, cost and complexity will be covered in section III. Some of the significant hardware impairments in terms of transceiver systems and individual components are discussed in section IV. In section V the Air interface design challenges are presented in the form of complexities in waveform, frame structure, channel coding, multiplexing and initial access design. Finally the conclusions and further work are presented in section VI.

## II. TRANSCEIVER ARCHITECTURAL OPTIONS AND CHALLENGES

The mm-wave mobile deployment framework is expected to be through small cells, as the propagation conditions at higher frequencies have direct impact on the coverage per cell. These small cells will likely perform as a heterogeneous network (HetNet) with larger Macro cells, provided either by the (then) legacy 4G or sub-6 GHz 5G systems. Also the mm-wave 5G mobile system will be likely to inter-work with other complementary RATs like WiGig and D2D (Device to Device). In this context the mobile handsets in particular will have to support multiple-RATs in the transceiver design, while ensuring compactness and low-power consumption. On the base station or access point more complex transceivers could be aimed, compared to the user equipment, providing higher beamforming capabilities.

### A. Challenges at the AP (Access Point) level

Generally base stations present higher design complexity due to the higher capabilities of the traditional base stations to meet the power consumption, cost and size challenges. However for the wider proliferations of mm-wave small cells, these challenges will have to be re-examined. Even considering that the small-cell AP would benefit from more relaxed specifications on these factors, much higher mm-wave performance specifications (transmit power, radiation gain, beamforming functions) will be placed on it. These specifications will have to overcome the issues related the link budget and fulfill requirements in terms of data rate, communication range, coverage and multi-user handling. As a consequence in mm-wave radio architectures the complexity is more asymmetrically focused on the base station (or AP) side [5].

At base station level, transceiver systems will be required to deal with a large number of antennas and hybrid beamforming capability, i.e. combining analog and digital beamforming. In this case, multi-module architectures will be favored in order to avoid the cost and reliability issues of large-size RFIC chips and packaged modules, and to provide a good scalability of the technical solution as a function of the small-cell specifications (range coverage). Antenna gain values, i.e. number of antenna elements, in these cases will be derived from link budget analysis based on the selected frequency band, the expected antenna gain at user equipment level, the expected communication range (small-cell radius), the signal-to-noise ratio, propagation conditions and outage statistics. For instance in radio access, if the AP is located between 2m and 6m, the beam steering capability of the antenna should allow to cover the whole sector of ±45° in azimuth, if four 90° sectors with independent modules are aimed, and about 80° in elevation, which would allow to cover the angle range from the AP to mobile users located in short distance cells.

A denser deployment of small cells would mean that a reconfigurable backhaul mesh network will be required for optimal management of the network capacity. Wireless backhaul or front-haul would be achieved through fixed point to point link in diverse frequency bands up to 86 GHz. Adaptive steerable links, possibly with multiple simultaneous beams, will be needed to enable dynamic backhaul link configurations, providing significant performance improvements, which results in high design complexities of the AP transceiver systems for mm-wave as well as for antennas. For these applications, transmit-array, reflect-array or lens antenna solutions could be aimed [6] but a technology gap in terms of active elements at higher frequency is needed to be filled to achieve the aimed configurability.

Finally the multiplicity of scenarios will require the network architecture to handle moving APs installed on board trains, cars, buses, etc. for high speed mobility use cases and this will have an impact on the beam steering capabilities and algorithms of backhaul nodes with very challenging latency issues.

*B. Challenges at the UE level*

Mobile users will benefit from a multi-RAT operation combining the new 5G standard in mm-wave band and legacy 4G LTE-A standard in the sub-6 GHz bands as well as other RATs such as WiFi and D2D. Current UEs are using multi-RAT transceiver chipsets with multi-band antennas, which are feasible because these standards are implemented in similar frequency bands and similar architectures. Generally the UE needs to be simple and transparent to the multi-antenna environment of the base station. The transceiver embedded in future mobile terminals, i.e. smartphone, tablets or handsets have strict requirements in terms of low power consumption (typically less than 1mW) and size (few cm² maximum). Such constraints, and particularly the low power consumption, have a direct impact on the limited degree of beamforming capabilities, even if the antenna sizes could fit in most of the terminals.

For UE equipment single-path T/R (Transmit/Receive) CMOS radio associated to a fixed small antenna arrays (or eventually a fixed beam antenna) could be aimed for a single and compact package [7]. On the other hand the angular coverage should be wide enough to guarantee robustness again in the UE device orientation. This could be obtained by including more than one module in the terminal and select the most favorable, exploiting limited beamforming capabilities.

Hence 5G mm-wave transceiver and antenna systems are expected to be in independent chipsets, due to their very different architecture and requirements for a close integration. This results in a technical challenge of densely packing multiple RF chains and antenna elements while guaranteeing their efficiency.

III. CHALLENGES IN HYBRID BEAMFORMING

By deploying antenna arrays at the transmitter and receiver, antenna gain can be exploited to compensate the adverse effects of propagation at mm-wave frequencies [8]. Since the size of the antenna elements decreases with the square of the frequency, the number of antenna elements that can fit in a given area can also increase at the same rate. Although this would allow to pack more antenna elements in a given physical antenna aperture, equipping each antenna with an RF (Radio Frequency) chain and high resolution Analog/ Digital Converters (DACs at the transmitter, ADCs at the receiver), as it is traditionally done in lower frequency systems, might be very costly and lead to high power consumption. This is in part due to the fact that high resolution converters with a high sampling rate (as required to process the large bandwidths expected for mm-wave systems) are costly and power hungry.

Several solutions have been proposed to address this constraint. On one hand, the transceivers can be implemented with low resolution converters [9, 10] On the other hand the number of RF chains and converters can be reduced by considering the hybrid beamforming architecture [11] which is depicted in Fig. 1. By performing part of the beamforming operations in the analog domain and the other part in the digital baseband, hybrid beamforming is able to provide a tradeoff between performance and system complexity/power consumption. For the analog pre-coder and combiner, the processing can be performed via a network of phase shifters or switches. In any case, enabling directional transmission with hybrid beamforming faces a series of challenges which are discussed below.

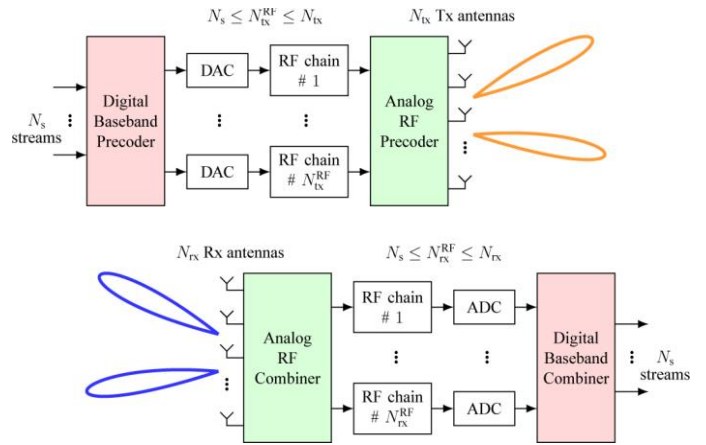

Figure 1: Transceiver Architecture for Hybrid Beamforming

One of the main challenges with hybrid beamforming is that the design of the pre-coder and the combiner. For the hybrid transceiver design, one needs to take into account that the hybrid pre-coder and combiner each consists of the product of a digital beamforming matrix and an analog beamforming matrix, which is subject to several limitations or constraints. For instance, when employing a network of phase shifters for the analog processing, the amplitude of the entries of the analog beamforming matrix is constant and only the phase of the entries can be modified. Moreover, it might not be possible in some cases to perform continuous phase shifting, such that only discrete phase shifting is available. Further possible limitations include the use of a limited set of code-words for the columns of the analog beamforming matrix. When taking into account such constraints on the analog processing, obtaining the optimum beamforming matrices, for instance to maximize the sum rate, is not a trivial task. Since the solution to such problems is unknown in general, one can instead recur to find approximations to the actual optimization problem [12, 13]. To this end, the sparse nature of the channel in angular domain can also be exploited.

Another important aspect to consider related to the analog processing and the large bandwidth expected for mm-wave transmissions, is the fact that the analog beamforming matrix is fixed for the entire frequency band. Although the digital component of the hybrid beamforming could be designed for each subcarrier, the analog beamforming matrix is frequency flat, i.e. it is constant across all subcarriers. This aspect needs to be taken into account for the transceiver design of multicarrier systems [14]. The same constraint arises similarly in a multiuser system, where the optimum set of analog beamformers needs to be considered across the given set of users [15].

For the previously discussed transceiver design, channel state information is required. However, estimating the actual channel might be a difficult task with the hybrid beamforming architecture, since the channel observed in the baseband can only be viewed through the constrained analog beamforming performed at both sides of the link. Thus, the constraints on the analog processing need to be considered when designing estimation procedures for acquiring channel state information. In addition, the transceiver design must also take into account that an estimate of the actual channel might not be available. In fact, the channel estimation and the transceiver design are to some extent coupled due to the hybrid architecture [16].

It is also important to highlight that attempting to estimate the channel matrix as traditionally done for smaller scale MIMO (Multiple Input Multiple Output) systems, would require a large training overhead due to the large number of antennas. On the other hand, the channel matrix at mm-wave frequencies is expected to have a sparse structure as a consequence of the reduced number of scattering clusters. Hence, the large channel matrices at mm-wave frequencies can be characterized with a reduced set of parameters including the angles of departures, the angles of arrivals and the path gains of each of the few paths. This sparse channel structure can be exploited to simplify the channel estimation procedure and reduce the training overhead [16].

One further inherent challenge of the hybrid architecture is the fact that the number of RF chains limits the number of streams that can be supported. Although this might not have a significant effect for single-user transmissions due to the expected low rank of the sparse channel matrices at higher frequencies, the number of RF chains does limit the number of users that can be served simultaneously.

IV. SPECIFIC HARDWARE IMPAIRMENTS AND MODELS

In this section we will look at specific impairment challenges of RF components at the mm-wave frequencies.

The losses and stray capacitances in RF components increase with carrier frequency. This manifests itself in decreasing efficiency and maximum power with increasing carrier frequency [17]. These two trends provide additional reasons for increasing the number antenna elements (in additional path-loss) and for applying one power amplifier per antenna element. Given the large number of antennas and high bandwidth on each antenna element it appears very unlikely that power amplifier linearization - operating typically at five times the RF bandwidth - is a sustainable model. Simpler pre-distortions schemes may need to be considered. We may also consider having a system which is limited by hardware impairments- e.g. a system where inter-carrier interference is a non-negligible source of interference.

A second major impairment source is phase-noise. Phase-noise arises in oscillators due to the impacts of active components and losses. Phase-noise causes changes in the phase of the signal which varies much faster than the phase of the propagation channel. In multi-carrier systems phase-noise also leads to leakage between the sub-carriers. This interference increases with the square of the carrier frequency and therefore becomes a major source of impairment in mm-wave systems. In multi-antenna systems a local oscillator signals (LO) will be required for each transmitter and receiver chain. Here three main solutions can be considered. The first would be to distribute the reference clock, typically 50-200MHz. and then use this to lock a voltage controlled oscillator using a phase-locked loop (PLL). A second solution is to generate a single LO signal and distribute this signal to multiple branches. Inside a chip, the second solution is attractive as it avoids the duplication of the PLL and VCO. However, when several receiver or transmitter chains are distributed across a circuit board a low cost substrate will be required which may drive cost and impose signal integrity and shielding issues. A third solution is to distribute a "scaled" LO i.e. a signal with a frequency a factor 2-12 of the desired LO which is then multiplied by a frequency multiplier or injection locked oscillator.

To capture and compare these different phase-noise solutions a phase-noise model as illustrated in Figure is used. The figure illustrates a linear-in-phase model where the noise added from the reference oscillator (Ref), phase-frequency detector and loop component (PLL) and voltage control oscillator (VCO) is indicated. In systems where the PLL and VCO is duplicated in every receiver chain – the noise from these components are independent while the noise from reference oscillator is identical. When the same LO is used – then all noise components are obviously identical. In [18] we presented a scenario with strong interference and showed that the single local oscillator gave the best result. In Figure 3 we consider a single-stream single-user case as a function of the number of antennas in transmitter and receiver (assumed equal). There is no noise and thus phase-noise is the only link impairment. Independent local oscillators are used in all chains. The performance with single local oscillator is identical to the result with single antenna. In this case separate local oscillators are most favorable.

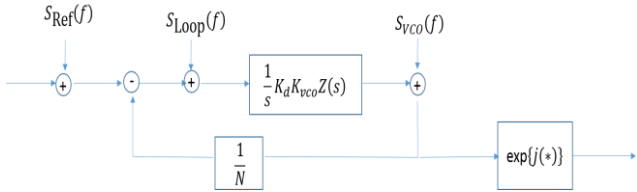

Figure 2: Phase-noise model

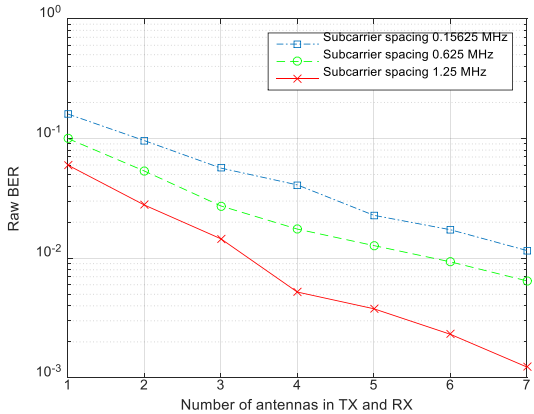

Figure 3: Raw BER with beamforming in the transmitter and receiver with 64QAM constellation. OFDM modulation over a 1GHz bandwidth, Phase-error compensation in each symbol.

V. CHALLENGES IN AIR INTERFACE DESIGN

The use of mm-wave technologies imposes specific challenges to the air-interface (AI), compared to sub-6 GHz frequencies. First, link budget constraints resulting from higher isotropic free-space loss lead to the need for higher antenna gains and the corresponding directional transmission. Directional transmission can change the effective channel characteristics and other system characteristics, e.g. interference characteristics leading to different requirements and design principles of AI development. Second, as observed in recent measurement campaigns [19], the number of path clusters as well as the angular spread of each cluster can be small. Moreover, reflection becomes dominant while refraction and diffraction become much weaker. When users are moving, mm-wave links suffer from blocking/shadowing as well as strong Doppler effects. Such different channel characteristics pose new challenges but also provide new opportunities for AI design, e.g. waveforms, frame structure, retransmission schemes etc.

We discussed the need for hybrid beamforming architectures in section III. Such architecture would have an impact on RF impairment modelling as well as on specific aspects of the AI, e.g. the initial access schemes. In addition, asymmetric antenna and RF configurations in uplink (UL) and downlink (DL) also affect AI design, e.g. by considering that UL coverage will be constrained by the much lower transmit power and beamforming gains expected at the user equipment.

The AI should support both standalone and non-standalone deployments, leading to challenges in initial access and control signaling. Accordingly, self-backhauling capabilities (where the same AI is used for both access and backhaul/fronthaul operation) may have to be considered in ultra-dense networks. The uncertainty in the finally released frequency bands for 5G mm-wave systems may demand different approaches for the above described challenges, thereby making AI design even more challenging.

The mm-wave AI design consists of five key technology components: waveforms, channel codes and re-transmission schemes, frame structure and numerology, multiple-access and duplexing schemes and initial access schemes. Taking into account the above challenges, these five key technology components should fulfill the following KPIs and design principles. More details are described in [20].

For the waveform, the following KPI's have been identified: Spectral efficiency, time-localization, PAPR (Peak to Average Power Ratio), MIMO compatibility, robustness against RF impairments (especially phase noise), complexity, flexibility, out-of-band emissions, robustness to frequency- and time selectivity of channels. The waveform should be designed taking into account mm-wave channel characteristics, multi-antenna techniques, transceiver characteristics, flexibility and scalability. A number of multi-carrier and single carrier waveforms have been identified as candidates, including CP-OFDM (Cyclic Prefix OFDM), Windowed OFDM, P-OFDM (Pulse-shaped OFDM), Unique-Word OFDM (UW-OFDM), Universal-Filtered OFDM (UF-OFDM), FBMC-QAM (Quadrature Amplitude Modulation) and FBMC-OQAM (Offset QAM). The single carrier candidate waveforms are DFT-s-OFDM (Discrete Fourier Transform-spread OFDM), ZT-DFT-s-OFDM (Zero-Tail DFT-s-OFDM), CPM-SC-FDMA (Continuous-Phase-Modulation Frequency Division Multiple Access) as well as D-QAM (Differential QAM) [20]. Such candidate waveforms will be assessed based on the above KPI's and using realistic channel and impairment models. A flexible and scalable waveform solution will be developed to cover wide range of scenarios, use cases and spectrum.

For channel code and re-transmission schemes, the main challenges are to fulfil very high throughput, reliability, and a wide range of latency requirements. The relevant KPI's include latency, throughput, throughput/chip area, error correction capability, complexity, and power/memory consumption. The capacity achieving codes such as Polar codes and spatially coupled LDPC are the promising candidates for mm-wave communication. Due to extremely high throughput, code complexity is a very important KPI. Therefore, low complex decoding techniques will be explored. Moreover, the necessity of enabling re-transmission should be assessed on a per use case basis. When needed, re-transmission schemes should fulfil KPI's in terms of latency, complexity, and reliability. The focus will be on a HARQ/ARQ design based on a combination of fast, efficient, and reliable feedback from the receiver. Furthermore, re-transmission mechanisms based on early error detection will be investigated. The designs of channel codes and re-transmission mechanisms will be kept mutually aligned.

For the frame structure and numerology, the wide contiguous bandwidths, high carrier frequencies, certain hardware impairments (e.g. phase noise) have to be taken into account, with the aim to fulfil high data rate and low latency. A large number of antenna elements need to be supported, calling for a scalable air interface. To allow enhanced energy efficiency, DTX and DRX should also be supported. Adaptability and re-configurability are important due to diverse deployment scenarios and different link types. Several different transceiver architectures should be supported seamlessly. For mm-wave AI, the most advanced frame structure candidates will be studied, evaluated and extended for mm-wave RAT, taking into account the mm-wave specific challenges. The aim is to achieve the user experience related targets, reduce complexity, enhanced energy efficiency, allowing for adaptability and re-configurability by developing a scalable frame structure that is agnostic to transceiver architecture. Fig. 4 shows a possible frame structure that may fulfil the above KPI's and design principles.

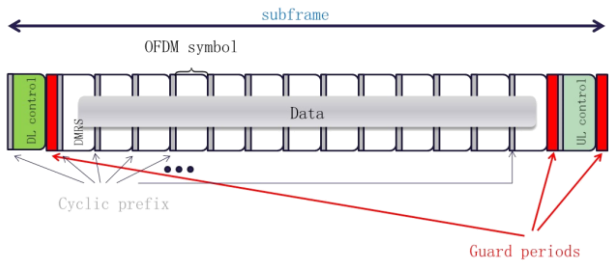

Figure 4: Possible frame structure design for mm-wave air-interface

The multiple access schemes should ensure the efficient use of the available communication resources (time, frequency, code and/or space), while duplexing schemes should allow dynamic matching of UL (Uplink) and DL (Downlink) traffic requirements. Both should take into account the extensive use of antenna arrays, support different antenna array configurations and transceiver architectures and imperfections in PHY implementation. The design should exploit context-aware information and relay selection to improve the network performance. The mm-wave AI multiple-access concepts will be developed for access and backhaul communication in different use cases and deployment scenarios, utilizing a mixture between SDMA and TDMA/FDMA/CDMA, exploiting context-aware algorithms and considering PHY imperfections. Analysis of TDD/ FDD (Time/ Frequency Division Duplex) schemes will be carried out taking into account massive arrays and high UL/DL traffic dynamics.

For initial access schemes, the key KPI's include the access delay (between user request and completion of initial access) and the overall communication overhead. Both stand-alone and overlay mm-wave networks should be considered. PHY imperfections play an important role in the design. Accordingly, two strategies can be applied: i) Consider some imperfections in the design; ii) Consider a design transparent to such imperfections. Furthermore, initial access should be coupled with beamforming and should exploit contextual information of UE's. For mm-wave AI, initial access mechanism will be designed considering highly directional transmissions, mobility, practical uplink constraints, existence of multiple neighboring AP's, sub-6 GHz macro coverage and different options of transceiver architectures, making use of advanced compressed sensing techniques and context-learning algorithms. Fig. 5 illustrates a possible scheme that leverages the sub-6 GHz macro coverage.

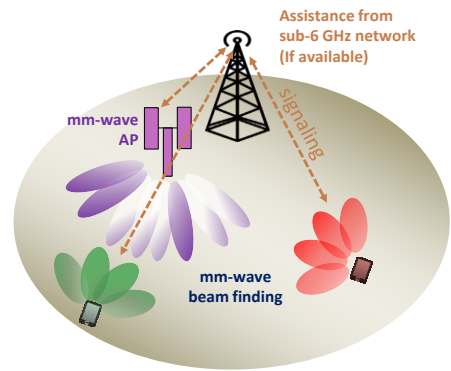

Figure 5: Initial access scheme

## VI. CONCLUSIONS AND FURTHER WORK

This paper explores the challenges and complexities in designing transceiver architectures and air interfaces for the 5G systems in mm-wave spectrum. There are multiple constraints brought on by the need for highly directional beamforming required at both ends for mm-wave communication links on transceiver design. The transceiver architectural options, to a large extent, will be governed by the need to keep the handsets cost, size and power efficient and to asymmetrically push the complexities to the AP side.

We analyze the hybrid beamforming architecture in detail and examine specific challenges within. One of the main challenges identified is that a large number of ADC/DAC units are required and running them at high sampling rate and at high resolution would be excessive in power consumption. Low resolution (even 1 bit) ADC/DAC design would thus need to be a priority. The combination of analog beamforming with the baseband digital beamforming also brings about many challenges like the analog component (based only on phase shifting) lacking precision and having to apply single analog phase shifting vectors across very wide bandwidths.

The hardware imperfections in these frequencies are also examined and with some initial models provided for signal corruption with phase noise.

The specific challenges in air interface design are presented in the form of five key technology components: waveforms, channel codes and re-transmission schemes, frame structure and numerology, multiple-access and duplexing schemes and initial access schemes. A possible frame structure for mm-wave operations is suggested, which includes features like scalability and enhanced energy efficiency.

Based on the challenges identified here, the mmMAGIC project will continue to develop robust schemes for mm-wave transceiver architectures and air interfaces. It is hoped that these solutions will significantly contribute to the pre-standardization discussions for mm-wave 5G systems in near future.

ACKNOWLEDGMENT

The research leading to these results received funding from the European Commission H2020 programme under grant agreement n°671650 (mmMAGIC project).